\documentclass[final,3p,times,twocolumn]{elsarticle}

\makeatletter
\def\ps@pprintTitle{
 \def\@oddfoot{}}
\makeatother

\usepackage{amssymb}
\usepackage{amsmath}

\usepackage{multirow}
\usepackage{colortbl}
\usepackage{xcolor}

\usepackage{url}

\journal{Computer Physics Communications}

\usepackage[per-mode=symbol,binary-units]{siunitx}
\DeclareSIUnit\dBm{dBm}
\DeclareSIUnit\dB{dB}
\DeclareSIUnit\baud{Baud}

\usepackage{listings}
\usepackage{placeins}

\begin{document}

\begin{frontmatter}



\title{Simulation of Nonlinear Signal Propagation in Multimode Fibers on Multi-GPU Systems}


\author[tudo]{Marius Brehler\corref{cor1}}
\cortext[cor1]{Corresponding author.}
\ead{marius.brehler@tu-dortmund.de}
\author[us]{Malte Schirwon}
\author[tudo]{Peter M. Krummrich}
\author[us]{Dominik G\"oddeke}

\address[tudo]{TU Dortmund, Chair for High Frequency Technology, 44227 Dortmund, Germany}
\address[us]{University of Stuttgart, Institute for Applied Analysis and Numerical Simulation, 70569 	Stuttgart, Germany}

\begin{abstract}
Mode-division multiplexing (MDM) is seen as a possible solution to satisfy the rising capacity demands of optical communication networks. To make MDM a success, fibers supporting the propagation of a huge number of modes are of interest. Many of the system aspects occurring during  the propagation can be evaluated by using appropriate models.
However, fibers are a nonlinear medium and, therefore, numerical simulations are required.
For a large number of modes, the simulation of the nonlinear signal propagation leads to new challenges,
for example regarding the required memory,
which we address with an implementation incorporating multiple GPU-accelerators.
Within this paper, we evaluate two different approaches to realize the communication between the GPUs and analyze the performance
for simulations involving up to 8 Tesla GPUs.
We show results for a MDM transmission system utilizing the extremely large but practically very relevant number of 120 spatial modes as an application example
and analyze the impact of the nonlinear effects on the transmitted signals.
\end{abstract}

\begin{keyword}
CUDA \sep
Fiber optics \sep
Multi-GPU \sep
Message passing interface \sep
Multimode fibers \sep
Space-division multiplexing \sep
Split-step Fourier method \sep
Fourth-Order Runge-Kutta\sep
Interaction Picture



\end{keyword}

\end{frontmatter}


\section{Introduction}
\label{sec:intro}

One of the main challenges in the design of future optical networks is to satisfy the growing capacity demand.
A very promising approach to solve this challenge is to use the yet untapped spatial dimension.
Space-division multiplexing (SDM) has attracted a lot of attention in the last years, both in industry and academic research.
One option to realize an SDM system is the use of multimode fibers (MMF), where each mode capable of propagation is used as
a channel for individual signals, referred to as mode-division multiplexing (MDM). \cite{richardson2013}\par
Recently, the utilization of 45 spatial modes in a multimode fiber as individual transmission channels was
demonstrated for the first time \cite{ryf2018_ECOC_2}.
With the availability of mode multiplexers for 45 Hermite-Gaussian modes~\cite{bade2018_OFC,fontaine2018_ECOC},
and to potentially excite even more modes~\cite{fontaine2018_OFC},
the investigation of MDM systems supporting a large mode count is getting more and more relevant.
During the design process of fiber optic transmission systems,
numerical simulations are the common choice to study different system aspects.
However, especially for a large mode count, new challenges arise within the simulation.

As fused silica is a nonlinear medium \cite{nlfo}, the simulation of light propagating in an optical fiber is quite challenging. 
The nonlinear signal propagation can be described by coupled partial differential equations for which a closed-form solution only exists in very few special cases.
Therefore, numerical methods are required to approximate solutions.
Exploring the impact of nonlinear effects in the case of data transmission,
is already challenging for only a single propagating mode, since long signal sequences need to be simulated.
Therefore, GPU-accelerators can be used to speed up simulations \cite{hellerbrand2010_OFC,pachnicke2010_ICTON,alcaraz2011_CPC}.
The numerical effort rises sharply when optical fibers which enable the propagation of multiple modes, especially fibers with a core diameter~\(\ge\SI{50}{\micro\meter}\), are the target of interest.
In those fibers, several tens/dozens 
or even more than 100 spatial modes can be used as spatial channels.
Moreover, the restricted amount of GPU-memory limits the approach to accelerate the simulation of the nonlinear signal propagation~\cite{uvarov2017_ICTON}. 
As a result, publications considering the nonlinear signal propagation in MMFs numerically are mostly limited to only a few modes if only a single GPU is used, e.g. 15 spatial modes in \cite{brehler2017_OFC}.
In this paper, we explore the possibility
to distribute the simulation of a transmission scenario in a single fiber
to multiple GPU-accelerators.
Here, we realize the communication between the GPUs with the Message Passing Interface~(MPI) or the NVIDIA Collective Communications Library (NCCL).
Only with multi-GPU implementations simulations with up to 36 spatial modes 
and 60 wavelength channels per mode \cite{brehler2018_OE} are possible,
for which a preliminary version of our MPI-implementation was used.

The paper is organized as follows: We first briefly present the mathematical description of the nonlinear signal propagation in multimode fibers.
Next, we review the numerical methods, followed by our MPI and NCCL implementations, as well as
GPU-specific modifications to the code required for the simulation of many modes.
The description of the implementation is followed by benchmarking the implementation incorporating up to 8 GPUs.
Finally, we use the application to demonstrate the simulation of an MDM transmission system in which a fiber with \SI{62.5}{\micro\meter} core diameter
provides~120 spatial modes as MDM channels.

\section{Modeling of the Nonlinear Signal Propagation in Multimode Fibers}

The nonlinear signal propagation in multimode fibers can be described by the nonlinear Schr\"odinger \cite{poletti2008} or the Manakov equation \cite{mumtaz2013_JLT,mecozzi2012_OE} for multimode fibers:

\begin{align}
\frac{{\partial {{\mathbf{A}}_\mathfrak{a}}}}{{\partial z}} =  \underbrace{- \frac{\alpha }{2}{{\mathbf{A}}_\mathfrak{a}} + i\sum\limits_{n = 0} {\left( {\frac{{{i^n}}}{{n!}}\, {{\beta _{n,\mathfrak{a}}}}\, \frac{{{\partial ^n}}}{{\partial {t^n}}}} \right)} {{\mathbf{A}}_\mathfrak{a}}}_{\hat{{L}}} \nonumber\\
+ \underbrace{i\gamma \left( {{\kappa _{\mathfrak{a}\mathfrak{a}}}{{\left| {{{\mathbf{A}}_\mathfrak{a}}} \right|}^2} + \sum\limits_{\mathfrak{b} \ne \mathfrak{a}} {{\kappa _{\mathfrak{a}\mathfrak{b}}}{{\left| {{{\mathbf{A}}_\mathfrak{b}}} \right|}^2}} } \right){{\mathbf{A}}_\mathfrak{a}}}_{\hat{{N}}}
\label{equ:nlse_manakov}
\end{align}
Here, \(\mathbf{A}\) represents the slowly varying envelopes of the spatial modes.
Within the linear part \(\hat{L}\),
the coefficient \(\alpha\) specifies the
attenuation, 
and the coefficients of the Taylor series expansion of the propagation constants are given by \(\beta_n\).
Within the nonlinear part \(\hat{N}\), the parameter \(\gamma\) is associated with the nonlinear refractive index change which is due to the Kerr-effect.
The intramodal nonlinear coupling coefficient is specified as~\(\kappa_{\mathfrak{aa}}\),
whereas the intermodal interaction is considered by~\(\kappa_{\mathfrak{ab}}\).
While only intramodal nonlinear effects occur during
the signal propagation in a single-mode fiber,
this is not the case for multimode fibers.
Here, the intermodal effects must be considered additionally.
Incorporating the weighted squared absolute values of \({\mathbf{A}}_\mathfrak{b}\) increases the numerical complexity significantly,
as discussed later.

For a more detailed description of modeling the nonlinear propagation in multimode fibers, see e.g. \cite{antonelli2016_JLT}.

\section{Numerical Approximation}

Since analytical solutions can only be calculated for a few special cases, numerical methods are required for the evaluation of Eq.~\eqref{equ:nlse_manakov}.
To approximate the solution of Eq.~\eqref{equ:nlse_manakov}, pseudo-spectral methods like the split-step Fourier method (SSFM) \cite{nlfo}
or the fourth-order Runge-Kutta in the Interaction Picture (RK4IP) method \cite{hult2007_JLT} can be used.

\subsection{The Symmetric Split-Step Fourier Method}
The formal solution to Eq.~\eqref{equ:nlse_manakov} is given by
\begin{align}
{A} (z+h,T) = \exp\left[h\left(\hat{L} + \hat{N}\right)\right] {A} (z,T) \,.
\end{align}
With the Baker-Hausdorff formula \cite{weiss1962_JMP}, this can be approximated as

\begin{align}
{A} (z+h,T) \approx \exp\left(h\hat{L}\right) \exp\left(h\hat{N}\right){A} (z,T) \,,
\end{align}
allowing to solve the linear part \(\hat{L}\) and the nonlinear part \(\hat{N}\) independently of each other.
The linear part \(\hat{L}\) is solved in the frequency domain and the nonlinear part \(\hat{N}\) is solved in the time domain.
The resulting splitting error can be further reduced by applying a symmetric split-step approach:

\begin{align}
{A} (z+h,T) \approx \exp\left(\tfrac{h}{2}\hat{L}\right) \exp\left(\int_z^{z + h}\hat{N}(z')\mathrm{d}z'\right) \nonumber\\\exp\left(\tfrac{h}{2}\hat{L}\right) {A} (z,T) 
\end{align}
The nonlinear part can be either solved by an iterative approach as described in \cite{nlfo} or with explicit schemes like the Runge-Kutta method.
This results in a third order accuracy to the step size~\(h\) and a global error of~\(O(h^2)\).
The two variants are denoted here as SSFM-Agrawal and SSFM-RK4, the latter using a fourth-order Runge-Kutta method to solve the nonlinear step.

\subsection{The Fourth-Order Runge-Kutta in the Interaction Picture Method}

To avoid the splitting error, Eq.~\eqref{equ:nlse_manakov} can be transformed with

\begin{equation}
A_I = \exp\left( -(z-z')\hat{L}\right)A
\label{equ:A_I}
\end{equation}
into the `Interaction Picture', where \(z'\) is the separation distance.
This allows to use explicit schemes like the fourth-order Runge-Kutta method, to solve  the differentiated form of Eq.~\eqref{equ:A_I}.
In contrast to the SSFM, no splitting is required, and the numerical accuracy is primarily limited by the applied explicit scheme.
With the separation distance \(z'\) defined as \(z + h/2\),
the algorithm to advance \(A(z,T)\) to \(A(z+h,T)\) is
\begin{subequations}\label{equ:algo}
\begin{align}
  {A_I} =& A\left( {z + \tfrac{h}{2},T} \right) = \exp \left( {\tfrac{h}{2}\hat{L}} \right) \cdot A\left( {z,T} \right) \hfill \label{equ:sub_AI}\\
  {k_1} =& \left( {\exp \left( {\tfrac{h}{2}\hat{L}} \right)\left[ {h\hat{N}\left( {A(z,T)} \right)} \right]} \right)A\left( {z,T} \right) \hfill\label{equ:sub_k1}\\
  {k_2} =& h\hat{N}\left( {{A_I} + \frac{{{k_1}}}{2}} \right) \cdot \left[ {{A_I} + \frac{{{k_1}}}{2}} \right] \hfill\label{equ:sub_k2}\\
  {k_3} =& h\hat{N}\left( {{A_I} + \frac{{{k_2}}}{2}} \right) \cdot \left[ {{A_I} + \frac{{{k_2}}}{2}} \right] \hfill\label{equ:sub_k3}\\
  {k_4} =& h\hat{N}\left[ {\exp \left( {\tfrac{h}{2}\hat{L}} \right) \cdot \left[ {{A_I} + {k_3}} \right]} \right] \nonumber \\
  &\cdot \exp \left( {\tfrac{h}{2}\hat{L}} \right) \cdot \left[ {{A_I} + {k_3}} \right] \hfill\label{equ:sub_k4}\\
  A\left( {z + h,T} \right) =& \exp \left( {\tfrac{h}{2}\hat{L}} \right) \cdot \left[ {{A_I} + \frac{{{k_1}}}{6} + \frac{{{k_2}}}{3} + \frac{{{k_3}}}{3}} \right] + \frac{{{k_4}}}{6}\hfill\label{equ:sub_A}
\end{align}
\end{subequations}
as given in \cite{hult2007_JLT}.
This method exhibits a local error of fifth-order and is globally fourth-order accurate.
A more detailed comparison between the SSFM and the RK4IP method, focusing on the nonlinear signal propagation in multimode fibers, is given in \cite{brehler2017_JLT}.

\section{Implementation of the Numerical Methods}

The signals can be represented by a matrix of sampled data with the dimension \(2M \times N\).
Here, \(M\) is the number of spatial modes, and the factor 2 results from taking both orthogonal polarization planes into account.
The number of discrete time samples is given by \(N\).
Thus, each row represents  a spatial or polarization mode.
To investigate Kerr-based nonlinear effects, namely self-phase modulation (SPM) and especially
cross-phase modulation (XPM), long symbol sequences need to be considered.
Furthermore, if wavelength-division multiplexing~(WDM) is of interest, and thus the impact of four-wave mixing (FWM) should be evaluated, each symbol needs to be represented by an appropriate number of samples
to simulate a sufficiently large frequency spectrum.
E.g. 256 samples per symbol, denoted as \(N_{sps}\), were used in \cite{brehler2017_OFC} to
simulate a spectral range of \SI{8.192}{\tera\hertz}.
In the referenced simulation, \(M=15\) spatial modes and \(N_\textrm{s} = 2^{14}\) symbols per spatial and per polarization mode were considered.
With \(N=N_s \cdot N_\textrm{sps}\), this results in a complex valued dense matrix of size \(30 \times 2^{22}\),
which requires \SI{1920}{\mebi\byte} of storage.
When further increasing the number of spatial modes~\(M\), the matrix containing the sampled signal might still fit into the GPU-memory, but not all intermediate results do any longer. We therefore propose to split the \(2M\) polarization and spatial modes to \(K\) processes.
Since \(N \gg 2M\), this approach has several advantages 
over splitting \(N\) contiguous samples of a unique spatial or polarization mode to different processes,
as discussed in the next section.

\subsection{Splitting the Numerical Problem}

\begin{figure}[h!]
 \centering
 \includegraphics[width=\columnwidth]{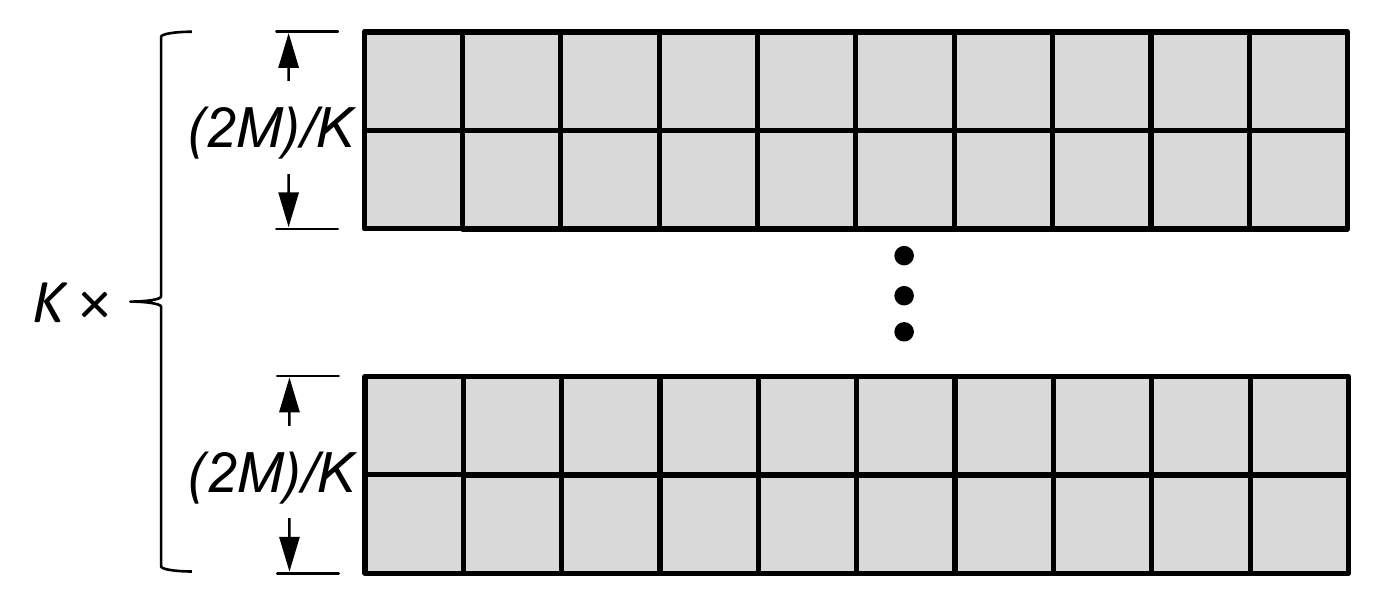}
 \caption{Signal matrix split to multiple processes.}
 \label{fig:matrix_split}
\end{figure}

In \cite{zoldi1999_SIAM}, \cite{taha2005_JS} the split-step Fourier method is parallelized by using distributed fast Fourier transform implementations.
However, this requires a lot of communication between the involved compute nodes.
Instead of letting multiple processes take part in the calculation of a spatial or polarization mode,
only entire modes are distributed to the different processes.
Here, each process is associated with one GPU, but the process itself can still involve multiple threads.
Thus, the \(N\) samples of a single signal are only required and processed by one unique process.
As proposed, the channels are equally distributed to \(K\) processes. With this, each process computes \(2M/K\) channels as illustrated in Fig. \ref{fig:matrix_split}

The matrix representing the sampled signal is stored row-major and the rows are aligned linear in the memory.
Therefore, the memory alignment is optimized for the fast Fourier transforms (FFT), as discussed in~\cite{brehler2017_JLT}.
The computation of the linear step \(\hat{L}\) can be executed fully parallel by each process independently.
Only for the calculation of the nonlinear step~\(\hat{N}\)
information from the other processes is required, namely the squared absolute values of the envelopes \({{{A}}}\) of all modes not locally available.

The SSFM-Agrawal requires the computation of \({{\left| {{{{A}}}} \right|}^2}\) once at the position~\(z\) for the first iteration and at the position~\(z+h\) for every following iteration.
Using the SSFM-RK4, the values \({{\left| {{{{A}}}} \right|}^2}\) are required to calculate \(k_1\), \(k_2\), \(k_3\), and \(k_4\), which is the same for the RK4IP method.
The squared absolute values \({{\left| {{{{A}}}} \right|}^2}\) can be stored real-valued.
Therefore, in every iteration or rather the calculation of \(k_n\), \(\left(2M - 2M/K\right)\cdot N\) real valued numbers have to be provided by the other processes and each process has to share its \(\left(2M/K\right)\cdot N\)
computed values.
The squared absolute values \({{\left| {{{{A}}}} \right|}^2}\)
are exchanged via MPI
or~NCCL.
Due to the large signal matrices, one has to expect quite large messages
even if communication is kept minimal with our splitting approach.
For the previous example with matrix size \(30 \times 2^{22}\), sharing all squared absolute values would result in a message size of \(\SI{960}{\mebi\byte}\).


In the following, we apply our modifications to the RK4IP method. The RK4IP allows more than doubled step-sizes \(h\) in the simulation of MDM transmission systems, as shown in \cite{brehler2017_JLT}.
Hence, less data exchange is required for the RK4IP method.
Nevertheless, the presented approach can be applied in an identical fashion to the SSFM-Agrawal and the SSFM-RK4.

\subsection{MPI-Implementation}

One option to realize the
communication between the involved GPUs
is to use the
the Message Passing Interface \cite{mpi31}.
Using MPI
has the advantage, 
that the GPUs do not necessarily have to be placed in the same compute node.
Here, one MPI process per GPU is used. 
With the availability of CUDA-aware MPI \cite{nvblog-cudampi} implementations, the programmer does not have to stage the data in the
host memory, as the GPU buffers can be directly passed to MPI.

A naive approach to realize the communication via MPI is the use of collective operations like \textbf{MPI\_Bcast} or \textbf{MPI\_Allgather}.
However, these rely on blocking communication and CUDA-aware implementations that supporting non-blocking collectives are still under development.
Using non-blocking communication instead has the advantage to overlap communication and processing of the data.
Overlapping communication and computations is essential to hide communication costs and to obtain good scalability.
We therefore decided to explicitly exchange data via asynchronous, and therefore non-blocking send and receive operations, namely \textbf{MPI\_Isend} and \textbf{MPI\_Irecv}.
The program sequence is described in Listing~\ref{lst:mpi}.
\begin{lstlisting}[caption={Basic programm flow to compute \(\hat N\) incorporating MPI.},label=lst:mpi]
void send_sqrabs(const int rank,
  const int size, const REAL *sqrabs,..,
  const int num_elem, MPI_Request *send_req) {
  
  for(int rk=0;rk<rank;rk++)
    MPI_Isend(&sqrabs[..],num_elem,
      MPI_DOUBLE,..,MPI_COMM_WORLD,
      &send_req[rk]);
      
  for(int rk=rank+1;rk<size;rk++)
    MPI_Isend(&sqrabs[..],num_elem,
      MPI_DOUBLE,..,MPI_COMM_WORLD,
      &send_req[rk-1]);
}

void recv_sqrabs(const int rank,
  const int size, REAL *sqrabs,..,
  const int num_elem, MPI_Request *recv_req) {
  
  for(int rk=0;rk<rank;rk++)
    MPI_Irecv(&sqrabs[..],num_elem,
      MPI_DOUBLE,..,MPI_COMM_WORLD,
      &recv_req[rk]);

  for(int rk=rank+1;rk<size;rk++)
    MPI_Irecv(&sqrabs[..],num_elem,
      MPI_DOUBLE,..,MPI_COMM_WORLD,
      &recv_req[rk-1]);
}

calc_squareabs(..);
recv_sqrabs(..);
send_sqrabs(..);
calc_nonlinear_own(..);

all_done = 0;
while(all_done < size-1) {
  MPI_Waitany(size-1, recv_req, &rk_idx,..);
  calc_nonlinear_others(..,rk_idx,..);
  all_done++;
}

apply_nonlinear_all(..);
\end{lstlisting}
After the computation of \(|A|^2\) for the \((2M)/K\) modes persisting on the GPU, 
we initialize the data exchange operations.
The values are send via~\textbf{MPI\_Isend} in the \textit{send\_sqrabs()} function, and matching receive \textbf{MPI\_Irecv} commands are posted in the
\textit{recv\_sqrabs()} function.
As mentioned before, these operations are non-blocking and therefore both commands return immediately, even if the transfers are not finished.
Next, the CUDA kernel is launched to calculate the contribution to the nonlinear phase rotation of the modes that are persisting on the GPU. This is non-blocking again.
Afterwards, a blocking operation \textbf{MPI\_Waitany} is called, to wait until any of the \textbf{MPI\_Irecv} commands has finished and if, the contribution of the received~\(|A|^2\) values to the nonlinear phase rotation is calculated.
If all~\(|A|^2\) values of the~\(K-1\) other processes are received, and all contributions are taken into account, the nonlinear phase rotation is finally applied to the modes persisting on the GPU.

This approach scales perfectly if the time needed to receive the next data is shorter than the
time for the simultaneously performed computations.
In this case, the GPU does not have to wait for the next data, since these are received while the GPU is performing computations.
The first work package is always available on the GPU, since this is the calculation of~\textit{calc\_nonlinear\_own()} for which no data needs to be received.
However, the execution of~\textit{calc\_nonlinear\_others()} relies on data sent from the other processes.
In practice, the possible overlap strongly depends on the simulation set-up, i.e. the number of spatial modes \(M\) and samples \(N\), and is limited by the number of involved GPUs \(K\) as well as the interconnects between the GPUs.

%


\subsection{NCCL-Implementation}

A higher-level approach is to exchange data via the NVIDIA Collective Communications Library (NCCL).
NCCL supports multiple GPUs installed in a single node or across multiple nodes. 
The library provides topology-aware collective communication primitives and features multiple ring formations for high bus utilization.
Within NCCL, the collectives are implemented in a single kernel
and are therefore associated to a so-called CUDA stream \cite{cudahandbook}.
The NCCL calls return when the operation is enqueued to the specified stream
and the collective operation is executed asynchronously.
In our implementation, \textbf{ncclAllGather} is used to aggregate the data.
As depicted in Listing~\ref{lst:nccl},
we use different streams for the kernel launch within \textit{calc\_nonlinear\_own()}
and the remaining kernel calls to enable concurrent execution.
\begin{lstlisting}[caption={Basic programm flow to compute \(\hat N\) incorporating NCCL.},label=lst:nccl]
calc_squareabs(..);
ncclAllGather((const void*)&sqrabs[..],
              (void*)sqrabs, num_elem,
              ncclDouble,comm,stream_a);

calc_nonlinear_own(..,stream_b);
cudaStreamSynchronize(stream_b);

for(int rk_idx=0; rk_idx < size-1; rk_idx++)
  calc_nonlinear_others(..,rk_idx,..,stream_a);
  
apply_nonlinear_all(..,stream_a);
\end{lstlisting}

To enable the implementation to utilize multiple nodes, we use NCCL together with MPI.
Hence, each GPU is associated with an MPI process as before.
A common NCCL communicator spanning all processes, is initialized as described in \cite[Example 2: One Device per Process or Thread]{NCCL}.

\subsection{GPU-Acceleration}
\label{subsec:gpu}
The GPU-acceleration of the RK4IP implementation is described in \cite{brehler2017_JLT} and \cite{brehler2018_NP}.
However, further modifications to the GPU code of our implementation are required in addition to the previously described adaptions.

In the preceding single-node implementation, only a single CUDA kernel capturing the nonlinear effects was launched.
As shown before, this is now split up into an \textit{own} kernel, responsible for calculating the nonlinear phase rotation of the locally stored modes,
and an \textit{others} kernel, responsible for the calculation for the nonlinear phase rotation induced by the modes not locally available.
Thus,~\(K-1\) instances of the latter kernel have to be launched. The overall nonlinear phase rotation is stored in an additional array of size \((2M/K)\cdot N\). Both kernels incorporate so-called shared memory, to alleviate the penalty occurring due to column access to the memory \cite{brehler2017_JLT, brehler2018_NP}.
In contrast to the single-node implementation, applying the nonlinear phase rotation to the locally available modes now only requires row access instead of column access to the memory. Applying the nonlinear phase rotation is performed by an additional kernel, as already indicated in Listings~\ref{lst:mpi} and~\ref{lst:nccl}.

\begin{figure}[b!]
 \centering
 \includegraphics[scale=0.5]{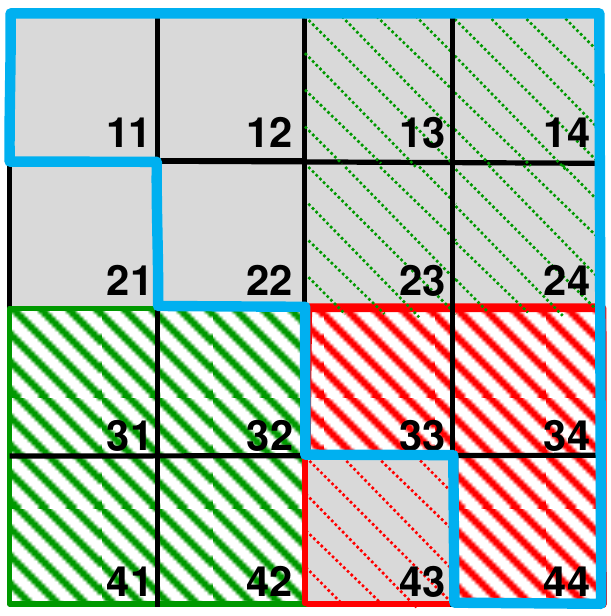}
 \caption{Exemplary interaction matrix for \(M=2\) and \(K=2\).
 The upper triangular matrix is highlighted in blue; All elements required for the calculations on GPU 2 are shaded red and green. By considering the symmetry, the light and dark shaded green elements are identical and only one of the sub-matrices needs to be stored.
 Furthermore, the light red shaded element can be neglected.}
 \label{fig:kappa}
\end{figure}

In addition, the interaction matrix no longer fits into the constant memory for a large number of modes without using a splitting approach.
Storing all \(\kappa\) values, requires a matrix of \(2M\times 2M\) elements. Assuming a symmetric matrix, which is the case for linearly polarized (LP) modes \cite{poletti2008},
it is sufficient to only store the upper triangular matrix, reducing the number of elements to \((2M \cdot 2M)/2 + M\).
This is exemplified for the case of \(M=2\) and \(K=2\) in~Fig.~\ref{fig:kappa}.
For a huge number of modes, e.g. \(M=120\), still~28920 double precision values of \SI{8}{\byte} would need to be stored in the constant memory, of which only \SI{64}{\kibi\byte} are available.
Therefore, this approach does not lead to a sufficient saving.
However, only~\(2M/K \cdot 2M \cdot 2 - (2M/K)^2\) need to be accessed for the calculations.
For GPU 2, these are the red and green shaded elements in Fig.~\ref{fig:kappa}.
The other elements of the matrix are only required on the other involved GPU.
Taking the symmetry into account again, it is sufficient to save only rows or columns which apply to the modes considered on the certain GPU.
With this in mind, the number of elements can be reduced to~\(2M/K \cdot 2M\).
Furthermore, for the \(\kappa\) coefficients describing the nonlinear coupling for the modes persisting in the GPU, it would be sufficient again
to only store the upper triangular matrix, as visualized in Fig.~\ref{fig:kappa}.
However, distributing the matrix via MPI and the necessary index arithmetic is more complicated for this case,
and only~\((2M/K \cdot 2M/K)/2 - M/K\) additional elements can be saved.

\section{Benchmark}
\label{sec:benchmark}

To achieve the maximum performance, peer-to-peer access between the GPUs is essential.
The benchmark is therefore performed on an AWS EC2-instance of type \textit{p2.8xlarge}.
This instance incorporates 4 Tesla K80 accelerators. Each K80 provides a pair of GK210 GPUs, resulting in~8 available GPUs. On this instance type the GPUs are connected via a common PCIe fabric.

The configuration used for the benchmark is given in Table~\ref{tab:set-up}.
\begin{table}[h!]
\centering
\caption{Configuration used for the benchmark.}
\label{tab:set-up}
{\scriptsize
\renewcommand{\arraystretch}{1.1}
 \begin{tabular}{c|c|c|c|c}
 \hline
  \(M\) & \(N_s\) & \(N_{sps}\) & \(K\) & \(M/K\) \\
  \hline
  \hline
  15  & \multirow{5}{*}{\(2^{14}\)} & \multirow{5}{*}{128} & 1 & \multirow{5}{*}{15} \\
  30  &                              &                       & 2 &                      \\
  60  &                              &                       & 4 &                      \\
  90  &                              &                       & 6 &                      \\
  120 &                              &                       & 8 &                      \\
  \hline
 \end{tabular}}
\end{table}
Considering a sequence with a symbol rate of \SI{32}{\giga\baud},
a spectral range of~\SI{4.096}{\tera\hertz} is simulated. 
Incorporating 8 GPUs allows to evaluate the nonlinear interaction between 120 spatial modes. This is of interest as it is the number of the potentially usable spatial modes in a fiber with~\SI{62.5}{\micro\meter} core diameter \cite{sillard2016_OFC}.

\begin{figure}[h!] 
 \centering
 \includegraphics[width=\columnwidth]{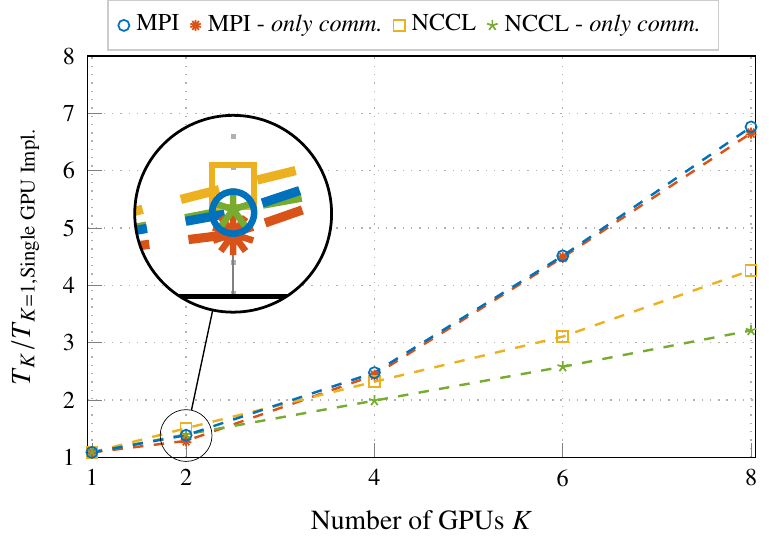}
 \caption{Scaling of the implementation.}
 \label{fig:scaling}
\end{figure}

The number of involved processes, or rather GPUs, is scaled up from 1 to 8, to investigate the scaling of the proposed implementations.
The number of spatial modes \(M\) persisting per GPU is kept constant.
In consequence, the total signal matrix occupies up to~\SI{7680}{\mebi\byte},
of which \SI{960}{\mebi\byte} are stored per GPU.
For every calculation of \(\hat{N}\), each GPU needs to share~\SI{480}{\mebi\byte}.
For the benchmark, 150 steps have been simulated and all calculations are executed with double precision.
Recall, that \(\hat{N}\) is calculated 4 times per step.
This is the same for an SSFM-RK4 implementation, whereas the number to calculate \(\hat{N}\) depends on the number of iterations in an SSFM-Agrawal implementation.
The initial distribution and the final collection of the sampled signal matrix, as well as the transfer of further necessary parameters and data, is excluded from the benchmark.
Results are shown in Fig.~\ref{fig:scaling}.
Here, the execution times \(T_K\) are normalized to the execution time of our previous single-node, single-GPU implementation \cite{brehler2017_JLT,brehler2018_NP}.

With only a single GPU used, \(K=1\),
the relative runtime is \(>1\).
Due to splitting the calculation of \(\hat{N}\) into several kernels,
the runtime increases by approximately \SI{8.5}{\percent}.
For \(K\le4\), the MPI- and NCCL-implementation scale nearly equal.
With even more GPUs involved,
the execution time of the MPI-implementation rises sharply. Incorporating all 8 GPUs, the MPI-implementation requires \(6.76\) times the execution time of the single-GPU implementation, whereas the NCCL-implementation scales with a factor of 4.26.
To evaluate the reason, the benchmarks are rerun without the additional calculations performed in the \textit{calc\_nonlinear\_others()}.
Therefore, only the amount of communication grows with an increasing \(K\).
Here, the MPI-implementation shows nearly identical results, whereas the relative runtime of NCCL-implementation drops.
For the MPI-implementation, this clarifies
that the increase of execution time is caused by communication, and not by the additional calculations.
\begin{table*}[htb!]
\centering
\caption{Mode groups (MG), \(A_\textrm{eff}\) in \si{\micro\meter\squared}, and number of spatial modes.} \label{tab:modes}
\resizebox{\linewidth}{!}{
{\scriptsize
\renewcommand{\arraystretch}{1.0}
\begin{tabular}{|c|llllllll|llllllll|c|}
\hline
MG&\multicolumn{8}{c|}{Modes in Group}&\multicolumn{2}{c}{\(A_\textrm{eff}\)}&\multicolumn{7}{|c|}{{\cellcolor[RGB]{192,192,192}}Total Num. of Spatial Modes up to this MG}\\
\hline
 1  &LP\(_{0,1}\) &&&&&&&&172 &&&&&&&&{\cellcolor[RGB]{192,192,192}}1\\
 2  &LP\(_{1,1}\) &&&&&&&&231 &&&&&&&&{\cellcolor[RGB]{192,192,192}}3\\
 3  &LP\(_{0,2}\) &LP\(_{2,1}\) &&&&&&&347 &311 &&&&&&&{\cellcolor[RGB]{192,192,192}}6\\
 4  &LP\(_{1,2}\) &LP\(_{3,1}\) &&&&&&&373 &372 &&&&&&&{\cellcolor[RGB]{192,192,192}}10\\
 5  &LP\(_{0,3}\) &LP\(_{2,2}\) &LP\(_{4,1}\) &&&&&&504 &469 &428 &&&&&&{\cellcolor[RGB]{192,192,192}}15\\
 6  &LP\(_{1,3}\) &LP\(_{3,2}\) &LP\(_{5,1}\) &&&&&&499 &545 &475 &&&&&&{\cellcolor[RGB]{192,192,192}}21\\
 7  &LP\(_{0,4}\) &LP\(_{2,3}\) &LP\(_{4,2}\) &LP\(_{6,1}\) &&&&&653 &605 &615 &521 &&&&&{\cellcolor[RGB]{192,192,192}}28\\
 8  &LP\(_{1,4}\) &LP\(_{3,3}\) &LP\(_{5,2}\) &LP\(_{7,1}\) &&&&&618 &690 &674 &561 &&&&&{\cellcolor[RGB]{192,192,192}}36\\
 9  &LP\(_{0,5}\) &LP\(_{2,4}\) &LP\(_{4,3}\) &LP\(_{6,2}\) &LP\(_{8,1}\)&&&&795 &732 &768 &733 &601&&&&{\cellcolor[RGB]{192,192,192}}45\\
 10 &LP\(_{1,5}\) &LP\(_{3,4}\) &LP\(_{5,3}\) &LP\(_{7,2}\) &LP\(_{9,1}\)&&&&731 &824 &835 &783 &635&&&&{\cellcolor[RGB]{192,192,192}}55\\
 11 &LP\(_{0,6}\) &LP\(_{2,5}\) &LP\(_{4,4}\) &LP\(_{6,3}\) &LP\(_{8,2}\)&LP\(_{10,1}\)&&&932 &853 &907 &900 &834 &671&&&{\cellcolor[RGB]{192,192,192}}66\\
 12 &LP\(_{1,6}\) &LP\(_{3,5}\) &LP\(_{5,4}\) &LP\(_{7,3}\) &LP\(_{9,2}\)&LP\(_{11,1}\)&&&841 &951 &979 &957 &879 &702&&&{\cellcolor[RGB]{192,192,192}}78\\
 13 &LP\(_{0,7}\) &LP\(_{2,6}\) &LP\(_{4,5}\) &LP\(_{6,4}\) &LP\(_{8,3}\)&LP\(_{10,2}\)&LP\(_{12,1}\)&&1066 &969 &1038 &1050 &1015 &926 &735 &&{\cellcolor[RGB]{192,192,192}}91\\
 14 &LP\(_{1,7}\) &LP\(_{3,6}\) &LP\(_{5,5}\) &LP\(_{7,4}\) &LP\(_{9,3}\)&LP\(_{11,2}\)&LP\(_{13,1}\)&&946 &1072 &1124 &1112 &1066 &966 &764 &&{\cellcolor[RGB]{192,192,192}}105\\
 15 &LP\(_{0,8}\) &LP\(_{2,7}\) &LP\(_{4,6}\) &LP\(_{6,5}\) &LP\(_{8,4}\)&LP\(_{10,3}\)&LP\(_{12,2}\)&LP\(_{14,1}\)&1195 &1081 &1162 &1188 &1174 &1119 &1009 &795 &{\cellcolor[RGB]{192,192,192}}120\\
 \hline
\end{tabular}}}
\end{table*}
In conclusion, communication and calculations can be perfectly overlapped using MPI,
additionally confirmed by profiling the application.
However, the implementation shows an improvable communication pattern for~\({K>4}\).
With the NCCL-implementation on the contrary, communication and calculations are not fully overlapping.
Anyway, the topology-aware communication patterns show clear benefits for the simulation with more than 4 GPUs involved.
For \(K=2\), highlighted in Fig.~\ref{fig:scaling}, the MPI-implementation is slightly outperforming the NCCL-implementation (factor 1.38 vs. 1.51).
With only two GPUs taking part in the simulation, the NCCL's topology-awareness cannot improve communication.
In this case overlapping of communication and calculations is much more important.

From a view point of weak scaling, an improved performance is desirable, especially for a large number of involved GPUs.
However, regarding the required all-to-all communication the performance metrics are not surprising. 
Nevertheless, the application enables the simulation of the nonlinear signal propagating of a huge number of spatial modes and a large frequency range, which was not possible so far.
Improved performance of the MPI implementation can be expected when decoupling the CPU-GPU control flow.
With the future availability of MPI-GDS \cite{venkatesh2017_mvapich},
the asynchronous send operations can be triggered directly after the squared absolute values are computed,
leading to better hiding of the communication.
In addition, also the optimization of collective operations is under investigation \cite{awan2016_EMPI,awan2018_EMPI}.
Therefore, future library implementations offer the potential to further improve the performance of the proposed implementation.
In the next section we show, what is already possible with the implementation based on the current available libraries.

\FloatBarrier
\section{Simulation of a \SI[detect-all=true]{62.5}{\micro\meter} Fiber}

Within the application example we demonstrate the feasibility of an MDM transmission over a multimode fiber with graded-index profile
featuring a core diameter of \SI{62.5}{\micro\meter} and a numerical aperture of 0.275.
The highest order modes feature small effective refractive indices
and are therefore affected most by the cladding.
In such a fiber, 120 of the spatial modes capable of propagation can be used for a mode-multiplexed transmission~\cite{sillard2016_OFC}.
We assume a profile exponent of~1.94 and a trench, a section with a reduced refractive index within the cladding of the fiber,
in a distance of \SI{1.25}{\micro\meter} to the core and with a width of \SI{3.5}{\micro\meter}.
The refractive index difference between the cladding and the trench is~\({6.5\cdot10^{-3}}\).
Further, an attenuation coefficient of~\SI{0.23}{\dB} is assumed for all modes.

The modes form strongly coupled groups of modes, meaning the linear coupling between the modes belonging to the same mode group is strong, whereas the linear coupling
between modes of different mode groups is weak. This is taken into account by choosing the appropriate nonlinear coupling coefficients \(\kappa\)
for the Manakov equation \cite{mecozzi2012_OE}.
The mode groups (MG) are considered as given in Table \ref{tab:modes},
which further provides the effective areas \(A_{\textrm{eff}}\) of the spatial modes.
The mode profiles and the later used propagation constants are calculated numerically with JCMsuite \cite{pomplun2007_PSSB}.
With the definition of~\(\gamma=(n_2\omega_0)/(c_0A_{\mathrm{eff,LP_{0,1}}})\) and a nonlinear refractive index \(n_{2}= \SI[per-mode=reciprocal,exponent-product=\cdot]{2.6e-20}{\meter\squared\per\watt}\), the fiber features a nonlinear parameter \(\gamma=\SI[per-mode=reciprocal]{0.61}{\per\watt\per\kilo\meter}\).
In the definition of~\(\gamma\), the center frequency of the optical signal is given by~\(\omega_0\) and~\(c_0\) is the speed of light in vacuum.
The differential mode group delays~\(\Delta\beta_1{=}\beta_{1,\mathfrak{a}} - \beta_{1,\textrm{LP}_{0,1}}\) are calculated
with Eq.~(29) from~\cite{antonelli2017_OE}. With \(\Delta\beta_1\) fulfilling the so-called phase-matching condition,
the cross-phase modulation between the strongly coupled groups is maximized. This is potentially the worst case for the nonlinear effects.
The phase matching condition for multimode fibers incorporates the group-velocity dispersion parameter \(\beta_2\), the second derivatives of the propagation constant \(\beta\), which are calculated numerically. The values assumed for the simulation for \(\Delta\beta_1\) and \(\beta_2\) are given in Table~\ref{tab:beta}.
\begin{table*}[htb!]
\centering
\caption{DMGDs \(\Delta\beta_1\) in \si{\pico\second\per\kilo\meter} and group-velocity dispersion parameters \(\beta_2\) in \si{\pico\second\squared\per\kilo\meter} for the different mode groups (MG).}
\label{tab:beta}
{\scriptsize
 \begin{tabular}{|c|ccccccccccccccc|}
  \hline
  MG & 1 & 2 & 3 & 4 & 5 & 6 & 7 & 8 & 9 & 10 & 11 & 12 & 13 & 14 & 15\\
  \hline
  \(\Delta \beta_1\) &   0.0 &  -7.32 &  -7.38 &  -7.48 &  -7.54 &  -7.60 &   -7.70 &  -7.79 &  -7.85 &  -7.95 &  -8.04 &  -8.18 &  -8.23 &  -8.39 &  -8.61 \\
  avg. \(\beta_2\)   & -23.1 & -23.3 & -23.5 & -23.8 & -24.0 & -24.2 & -24.5 & -24.8 & -25.0 & -25.3 & -25.6 & -25.9 & -26.2 & -26.7 & -27.4 \\
  \hline
 \end{tabular}}
\end{table*}

In the simulation, each spatial mode carries 60~WDM
channels within a~\SI{50}{\giga\hertz} grid.
The center frequency of the wavelength channel with the lowest carrier frequency is placed at~\SI{191.95}{\tera\hertz},
the one with the highest carrier frequency is placed at~\SI{194.9}{\tera\hertz}. 
Thus, a spectral bandwidth of~\SI{3}{\tera\hertz} 
is used for transmission.
Within the simulation, each WDM channel carries a dual polarization (DP) Quadrature Phase-Shift Keying (QPSK) modulated signal, with a symbol rate of~\SI{32}{\giga\baud}.
The average launch power per DP-QPSK signal is set to~\SI{-1}{\dBm}.
Here, we simulate the transmission over two~\SI{80}{\kilo\meter} spans,
resulting in a transmission distance of \SI{160}{\kilo\meter}.
After each span, the fiber losses are compensated by a noiseless amplifier with flat-gain profile.
White noise is added to the signals before the receiver,
setting the optical signal-to-noise ratio (OSNR) to \SI{20}{\dB}.
In this regime, the nonlinear effects are the dominating source of the signal degradation.
Within the digital signal processing stage, the dispersion is perfectly compensated.
Finally, a clock recovery \cite{schmogrow2012_OE} as well as a phase recovery are applied \cite{tsukamoto2006_PTL}.
To quantify the nonlinear impairments, the squared Q-factors are estimated for each mode and each WDM channel.

\begin{figure}[ht!]
 \centering
 \includegraphics[width=\columnwidth]{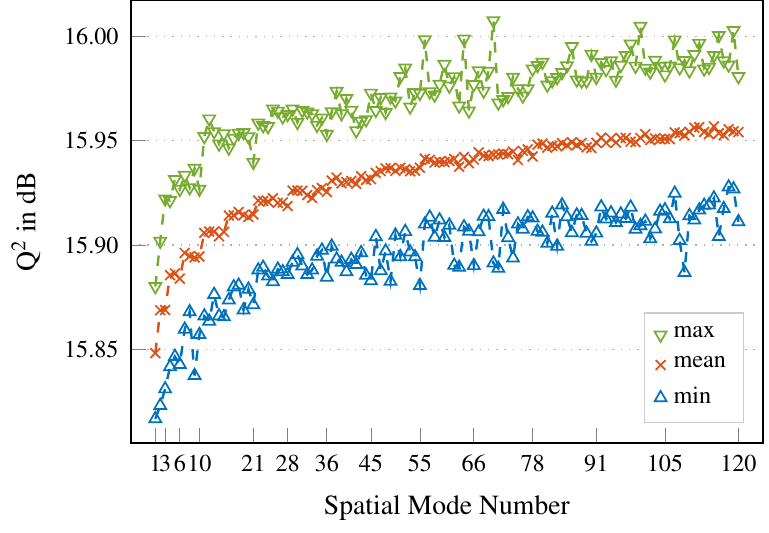}
 \caption{Squared Q-factors for each spatial mode after transmission over \SI{160}{\kilo\meter}, evaluated for an OSNR of \SI{20}{\dB}.}
 \label{fig:Q-160km}
\end{figure}

\begin{figure}[ht!]
 \centering
 \includegraphics[width=\columnwidth]{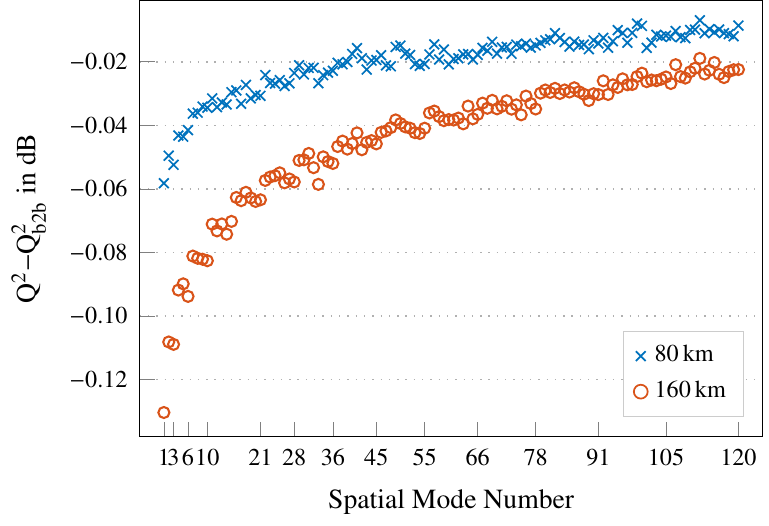}
 \caption{Mean squared Q-factors for each spatial mode after transmission over 80 and \SI{160}{\kilo\meter}, relative to the squared Q-factors obtained for a back-to-back (b2b) transmission.}
 \label{fig:Q-b2b}
\end{figure}

The minimal, mean, and maximal Q\(^2\)-factors for each mode after the transmission over \SI{160}{\kilo\meter} are depicted in Fig.~\ref{fig:Q-160km}.
Since the fundamental mode features the smallest effective areas \(A_{\mathrm{eff}}\)
and the highest coupling coefficients,
it suffers most from the nonlinear impairments and features the smallest Q\(^2\)-factor.
For the higher order modes, the mean Q\(^2\)-factors improve.
To assess the nonlinear signal distribution, we evaluate the mean Q\(^2\)-factors relative to Q-factors obtained for a back-to-back (b2b) transmission,
shown in Fig.~\ref{fig:Q-b2b}.
With about \SI{-0.06}{\dB} after the first and \SI{-0.13}{\dB} after the second \SI{80}{\kilo\meter} span,
the fundamental mode again shows the highest signal degradation.
Independent of the transmission distance, the higher order modes are less affected by the nonlinear effects.
However, one can clearly identify the lower order mode groups,
especially based on the results for a transmission over \SI{160}{\kilo\meter}.
Also the induced nonlinear penalty increases with increasing transmission distance,
the overall penalty is rather small.
For lower OSNRs, an even smaller penalty can be expected~\cite{brehler2018_OE}.
Hence, the utilization of 120 spatial modes in an MDM transmission system seems possible,
and the Kerr-based nonlinear impairments do not prohibit the use of such a fiber.

\FloatBarrier
\section{Conclusion}
In this paper,
we presented a multi-GPU implementation to simulate the nonlinear signal propagation in multimode fibers.
This allows the simulation of a huge number of spatial modes while considering a large spectral bandwidth at the same time.
We revealed necessary modification in order to simulate many spatial modes
and discussed various approaches how to realize the communication between the GPUs.
The performance of the implementation was analyzed, whereas the communication between the GPUs was realized with either MPI or NCCL.
While MPI shows performance benefits for a few used GPUs,
the implementation clearly profits from NCCL's topology-awareness if more than 4 GPUs are involved in the simulation.
For the first time,
it was possible to simulate a mode-division multiplexing system utilizing 120 spatial modes in a \SI{62.5}{\micro\meter} fiber
along with 60 wavelength channels per spatial mode
by using 8 GPUs.
In the application example, we have evaluated the nonlinear impairments for each spatial mode and each wavelength channel.
The results allow to conclude that the nonlinear impairments do not prohibit the usage of such a large number of spatial modes
in a mode-division multiplexing system.
Regarding the nonlinear effects, it can be expected that one can scale up mode-division multiplexing systems far beyond the most recent transmission experiment in which 45 spatial modes were transmitted over \SI{26.5}{\kilo\meter} \cite{ryf2018_ECOC_2}.
The implementation presented here allows to study those future systems.


\section*{Acknowledgments}
The authors are grateful for the donation of a Tesla K40c by NVIDIA
through the GPU Grant program.
The work of M. Schirwon and D. G\"oddeke was supported by the German Excellence Initiative through EXC 310~(SimTech).





\end{document}